\documentclass[runningheads]{llncs}
\usepackage{graphicx}
\usepackage{booktabs}
\usepackage{hyperref}
\usepackage{amsmath}
\usepackage{orcidlink}

\begin{document}
\title{Fast 3D registration with accurate optimisation and little learning for Learn2Reg 2021}
\titlerunning{Fast 3D registration with accurate optimisation and little learning}
%
\author{Hanna Siebert\orcidlink{0000-0003-1651-6960} \and
Lasse Hansen\orcidlink{ 0000-0003-3963-7052} \and
Mattias P. Heinrich\orcidlink{0000-0002-7489-1972}}

\authorrunning{H. Siebert and L. Hansen and M. P. Heinrich}
%
\institute{
Institute of Medical Informatics, Universität zu Lübeck, Germany\\
\email{\{siebert,hansen,heinrich\}@imi.uni-luebeck.de}}
\maketitle              
\begin{abstract}
Current approaches for deformable medical image registration often struggle to fulfill all of the following criteria: versatile applicability, small computation or training times, and the being able to estimate large deformations. Furthermore, end-to-end networks for supervised training of registration often become overly complex and difficult to train. For the Learn2Reg2021 challenge, we aim to address these issues by decoupling feature learning and geometric alignment. First, we introduce a new very fast and accurate optimisation method. By using discretised displacements and a coupled convex optimisation procedure, we are able to robustly cope with large deformations. With the help of an Adam-based instance optimisation, we achieve very accurate registration performances and by using regularisation, we obtain smooth and plausible deformation fields. Second, to be versatile for different registration tasks, we extract hand-crafted features that are modality and contrast invariant and complement them with semantic features from a task-specific segmentation U-Net. With our results we were able to achieve the overall Learn2Reg2021 challenge's second place, winning Task 1 and being second and third in the other two tasks.

\end{abstract}
\section{Motivation}
Deep-learning-based approaches for medical image registration usually involve an elaborate learning procedure and yet they often struggle with the estimation of large deformations and the versatile usability for a wide range of tasks. To address the different registration tasks of the Learn2Reg2021 challenge\footnote{\href{https://learn2reg.grand-challenge.org}{https://learn2reg.grand-challenge.org}}, we present a fast and accurate optimisation method for image registration that requires little learning. Our method robustly captures large deformations by using discretised displacements and a coupled convex optimisation. In order to be versatile for various tasks, we include a hand-crafted feature extractor in our method that is contrast and modality invariant and still highly discriminative for local geometry.

\section{Methods}
The main idea of our method is to perform large-deformation image registration by using a coupled convex optimisation \cite{heinrich2014non} that approximates a globally optimal solution of a discretised cost function followed by an Adam-based instance optimisation to further improve the local registration accuracy. Dense correlation has already been used extensively in learning based optical flow estimation (cf. PWC-Net \cite{sun2018pwc}) and end-to-end trainable 3D registration networks (cf. PDD-Net \cite{heinrich2019closing}), however both approaches have limitations. PWC-Net requires multiple warping steps and is difficult to extend from 2D to 3D (see \cite{gunnarsson2020learning}). PDD-Net employs a dense 3D displacements, but substantially simplifies the optimisation strategy, which may lead to some inaccuracies. ConvexAdam aims to combine the best of both worlds (learning and optimisation-based) by leveraging segmentation priors where available and relying on robust hand-crafted features and fast discrete optimisation.   

\begin{figure}
\includegraphics[width=\textwidth]{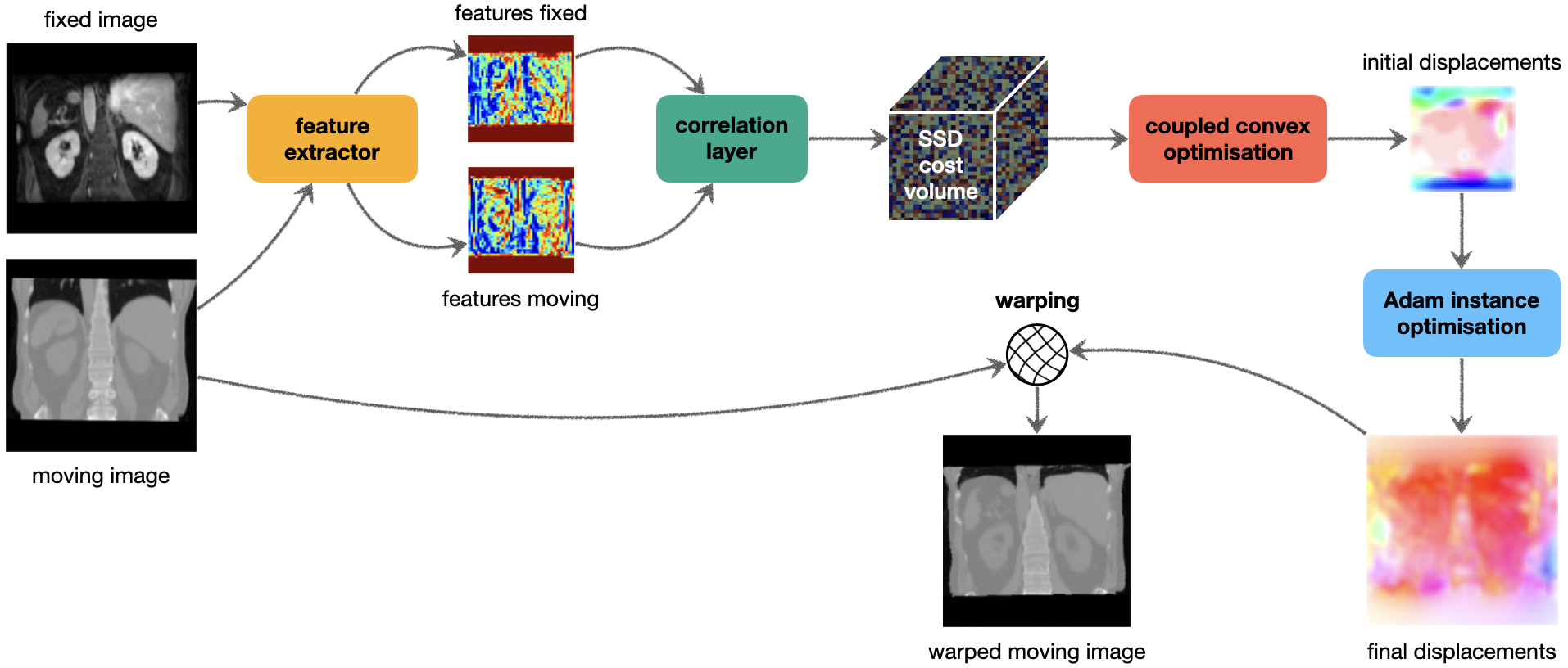}
\caption{The structure of our registration method. It consists of a feature extractor (MIND and/or nnUNet) and a dense correlation layer followed by a coupled convex optimisation and an Adam-based instance optimisation.} \label{fig_method}
\end{figure}

As visualised in Figure~\ref{fig_method}, the basic structure of our registration method consists of a feature extractor, a correlation layer, a coupled convex optimisation, and an instance optimisation. 

The feature extractor outputs contrast and modality invariant features from the fixed and moving input images. For this, hand-crafted MIND features \cite{heinrich2013towards} ensuring versatility regarding different types of registration tasks can be employed. Depending on the availability of labelled image data, automatic segmentations as provided by the nnU-Net \cite{isensee2021nnu} can be used instead. Different to other state-of-the-art supervised deep learning registration methods \cite{mok2020large} we avoid using the expert labels only at the end for the warping loss, which may lead to sub-optimal results due to limited gradient backflow. We instead found that using off-the-shelf segmentation networks produce best results.

The obtained features are fed into a correlation layer, which computes a sum-of-squared-differences (SSD) cost volume with a box filter and gives an initial best displacement for each voxel (simply taking the $\operatorname{argmin}$). Therefore, we employ a search space with up to $5000$ discretised displacements per voxel. The capture range can be up to at least 48 voxels in each dimension (setting for Task 2) and therefore estimate large motion accurately.

The correlation layer's output is used to solve two coupled convex optimisation problems for efficient global regularisation: In several iterations, alternating steps are performed for similarity and smoothness optimisation, i.e. a spatially smoothed field based on the current $\operatorname{argmin}$ (minimal SSD costs) displacements followed the by adding a penalty to the discreted SSD costs based on the discrepancy of this current globally smooth optimum. 

The resulting displacements in turn are used as a starting point for an Adam-based instance optimisation in order to provide the final deformation grid used for warping of the moving input image. This step is very similar to classic optical flow estimation \cite{papenberg2006highly}. For this purpose, the cost function is linearised and the Adam optimiser \cite{kingma2014adam} is used for gradient descent. Smoothness of the displacement field is induced by adding a B-spline deformation model and diffusion regularisation.

\section{Experiments and Results}

\begin{figure}
\centering
 
  \begin{tabular}{ccc}
     \includegraphics[height=0.25\linewidth,trim={1.2cm 0 0 0},clip]{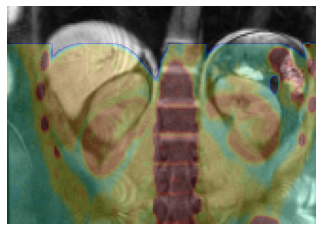}&
     \includegraphics[height=0.25\linewidth,trim={0 0.5cm 0 0},clip]{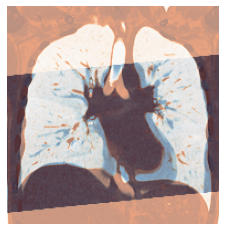}&
     \includegraphics[height=0.25\linewidth]{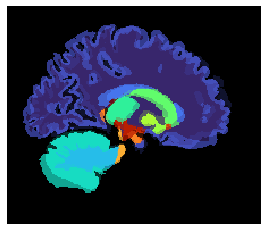}\\
     \includegraphics[height=0.25\linewidth,trim={1.2cm 0 0 0},clip]{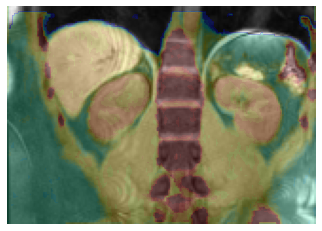}&
     \includegraphics[height=0.25\linewidth,trim={0 0.5cm 0 0},clip]{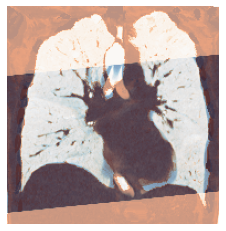}&
     \includegraphics[height=0.25\linewidth]{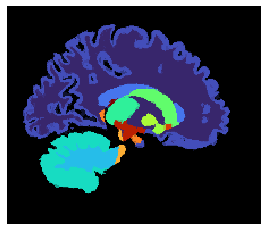}\\
     Task 1& Task 2 & Task 3 \\
  \end{tabular}
  
  \caption{Qualitative results of our proposed method (top row: colourmap overlay of fixed and moving image (Task 1 and 2) or segmentation (Task 3); bottom row: overlay of fixed and warped moving image or segmentation).}
  \label{quantitative_results}
\end{figure}

Each of the Learn2Reg2021 tasks entails certain challenges that we face with slightly varying experimental setups as outlined in the following. The complete implementation details can be found in our publicly available repository\footnote{\href{https://github.com/multimodallearning/convexAdam}{https://github.com/multimodallearning/convexAdam}}. Table~\ref{tab1} presents quantitative results and Figure~\ref{quantitative_results} shows qualitative results for the individual tasks.

\begin{table}
\caption{Results for the different Learn2Reg2021 tasks. Accuracy is measured by the Dice similarity of organ segmentations (Dice), the target registration error for anatomical landmarks (TRE), and the $95\%$~Hausdorff distance for segmentations (HD). Robustness is measured by the $30\%$ lowest Dice scores ($\mathrm{Dice}_{30}$), Dice scores for additional segmentations ($\mathrm{Dice}_{+add}$) and the $30\%$ highest TRE values ($\mathrm{TRE_{30}}$). Plausibility of the deformations is measured by the standard deviation of the logarithmic Jacobian determinant (SDlogJ). Dice similarities are reported in $\%$, TRE and HD values are given in millimetres and inference time is given in seconds. The last table displays the challenge scores and ranks for the overall 1st, 2nd, and 3rd place.}\label{tab1}

\begin{centering}
 \scriptsize{
\begin{tabular}{ccc}

\begin{tabular}{r|c|c|c|c|c}
\toprule
\multicolumn{6}{c}{\small\textbf{Task 1}}\\
\midrule
 &  $\mathrm{Dice}$ & $\mathrm{Dice}_{+9}$ & $\mathrm{HD}$ & $\mathrm{SDlogJ}$ & $\mathrm{time}$\\
  \midrule
 initial&$33.1$ & $22.3$ & $ 44.48 $ & $-$ & $-$ \\
  \midrule
 ours&$75.4$&$73.1$& $20.75$ & $0.09$ & $1.30$ \\
\bottomrule
\end{tabular}

 &$~~~~~~$&
 
\begin{tabular}{r|c|c|c|c}
\toprule
\multicolumn{5}{c}{\small\textbf{Task 2}}\\
\midrule
 &  $\mathrm{TRE}$ & $\mathrm{TRE}_{30}$ & $\mathrm{SDlogJ}$ & $\mathrm{time}$\\
  \midrule
 initial & $10.24$ & $16.80$ &-&-\\
  \midrule
 ours&$1.85$ & $2.89$ & $0.06$ & $1.82$\\
\bottomrule
\end{tabular}

\\
$~$
\\
$~$
\\
\begin{tabular}{r|c|c|c|c|c}
\toprule
\multicolumn{6}{c}{\small\textbf{Task 3}}\\
\midrule
 &  $\mathrm{Dice}$ & $\mathrm{Dice}_{30}$ & $\mathrm{HD}$ & $\mathrm{SDlogJ}$ & $\mathrm{time}$\\
  \midrule
 initial & $55.9$ & $29.7$ & $4.07$ &-&-\\
  \midrule
 ours&$79.9$ & $64.5$ & $2.00$ & $0.05$ & $12.62$\\
\bottomrule
\end{tabular}

 &&

 \begin{tabular}{r|c|c|c}
\toprule
\multicolumn{4}{c}{\small\textbf{Scores and ranks}}\\
\midrule
 & {\textbf{Task1}}
 & {\textbf{Task2}}
 & {\textbf{Task3}}\\
 
 \midrule
 
 & score (rank) & score (rank) & score (rank) \\
  \midrule
  
 LapIRN & $0.86$ ($2$) & $0.79$ ($4$) & $0.94$ ($1$)  \\
   \midrule
  convexAdam & $0.88$ ($1$) & $0.83$ ($3$) & $0.82$ ($2$)   \\
  \midrule
   PIMed & $0.85$ ($4$) & $0.68$ ($6$) & $0.70$ ($5$)   \\
\bottomrule
\end{tabular}

\end{tabular}
}
\end{centering}

\end{table}

\paragraph{Task 1 thorax-abdomen CT-MR.}
The first task aims to align multimodal intra-patient data. Besides of multimodal image registration, the objectives of learning from few and noisy labels, as well as dealing with large deformations and missing correspondences are challenging.
For this task, we extract hand-crafted MIND features and include an inverse-consistency constraint as introduced in \cite{heinrich2014non} to enforce a minimised discrepancy between the forward and backward transformations in order to avoid implausible deformations. To further regularise the displacement field during Adam instance optimisation, we add thin plate splines yielding smooth deformation fields. As large deformations are to be expected, we chose a search space that includes discretised displacements with a capture range of $64~\mathrm{mm}$ for each dimension within the scanned anatomy.

\paragraph{Task 2 lung CT.} 
The second task is to perform inspiration-expiration registration on intra-patient lung CT data. In this task, there is the challenge of estimating large breathing motion for scans with only partial visibility of the lungs in the expiration scans. The displacement search range is selected in order to capture motion with up to $42\times~30~\times~42~\mathrm{mm}$ for the $x$-, $y$-, and $z$-dimension respectively. Like in the first task, MIND features of both input images are used to compute the SSD cost volume.

\paragraph{Task 3 whole brain MR.}  
The third task deals with the registration of inter-patient T1-weighted brain MRI. Here, the main challenge is to precisely align small structures of variable shape. For this reason, we chose a displacement capture range of $16~\mathrm{mm}$ for each dimension within the scanned brain structures. As this task comprises a large amount of labelled image data, nnU-Net predictions for segmentation guidance are employed. We use the nnU-Net predictions in the form of inverse class-weighted one-hot encodings as features for our method's optimisation steps.

\section{Conclusion}
Our contribution to the Learn2Reg2021 challenge showed that image registration can be performed fast and accurately using an optimisation strategy with little learning. It is highly parallelisable on a GPU and robust by using a large search space of discretised displacements. Smoothness of the deformation fields could be induced by a global convex regularisation, diffusion regularisation, and B-spline interpolation. By using an efficient Adam-based instance optimisation, our method yields very precise results and by integrating a modality-invariant feature extractor, we achieve a wide versatility. We were able to achieve the overall Learn2Reg2021 challenge's second place, winning Task~1, being second in Task~3, and being third in Task~2.

%
%
%
%
\bibliographystyle{splncs04}

\end{document}